Article

# Psychophysical Interpretation of Quantum Theory

**Rajat K. Pradhan**

**ABSTRACT**

It is shown that the formalism of quantum theory naturally incorporates the psychophysical parallelism and thereby interprets itself, if the subjective aspects are taken as equal partners alongside the objective aspects as determinants of *Reality as a Whole*. The inevitable interplay of the subject (observer) and the object (observed) in making up Reality is brought out succinctly through a comprehensive psychophysical interpretation which includes in its bosom the truths of many of the major interpretations proposed so far as essential ingredients. At the heart of this novel approach lies the interpretation of the complex conjugate quantities such as the conjugate wave function Ψ*(**r**, t), the bra vector <Ψ|, and the adjoint operator **A**†etc. as representing the subjective counterparts of the corresponding objective aspects represented by the wave function Ψ(**r**, t), the ket vector |Ψ>, and the observable **A** etc. respectively. This brings out the psycho-physical parallelism lying hidden in the quantum mechanical formalism in a quite straightforward manner. The measurement process is shown to be a two-step process comprising objective interaction through the retarded waves and subjective observation leading to rise of knowledge through the advanced waves.

## 1. Introduction

A century after the advent of quantum theory and in spite of the unenviable success it has achieved in explaining diverse phenomena ranging from the microscopic elementary particles to the macroscopic universe itself, it suffers from a serious deficiency which is denoted by the general phrase *the interpretation problem* (Albert, 1992; d'Espagnat, 1976; d'Espagnat, 1979; Home, 1997). While the QM formalism offers readymade handy tools for calculation purposes, understanding the meaning of the wave function, its collapse in the measurement process to an *eigenstate* and nonlocal correlations among spatially separated components have been extremely difficult issues ever since its inception. The difficulty was very well paraphrased by an exasperated Feynman (1967) when he remarked "*I think it is safe to say that no one understands quantum mechanics*" and this, notwithstanding the fact that he happens to be the one who developed the path integral approach to quantum mechanics.

Interpretations have been proposed mainly along two distinct lines- those which avoid or deny the necessity of a conscious observer and those which admit of such a necessity.

The first category starts with the original Born (1926) interpretation known as the ensemble/statistical interpretation which was supported by Einstein (Karanth, 2011) and later on espoused by Ballentine (1970); then, there is the official, textbook interpretation known popularly as the Copenhagen interpretation enunciated and elucidated by Bohr (Bohr, 1935) and Heisenberg (Heisenberg, 1927); then comes the pilot wave interpretation of de Broglie (de Broglie, 1927; de Broglie, 1925) ,which was later on modified by Bohm (1952) to the well-known de Broglie-Bohm hidden variables theory; the Quantum logic scheme of Birkoff and von Neumann (1936), the original relative state formulation of Hugh Everett (Everett, 1957) expanded and popularized by deWitt and Graham as the many Worlds interpretation (deWitt and Graham, 1973); Nelson's Stochastic quantum mechanics (Nelson, 1966) in which the Schrödinger equation emerges as a consequence of a Markov process; the objective collapse theories (Ghirardi, Rimini and Weber, 1986; Bassi and Ghirardi, 2003) modifying the Schrödinger equation by adding non-linear terms to bring about spontaneous wave function collapse; the Histories approach (consistent and decoherent) of Omnes (1994), Griffiths (Griffiths, 1984; 2002), Gell-Mann and Hartle (1991; 1993), the strikingly straightforward transactional interpretation of Cramer (Cramer, 1986; 1988) based on the Feynman-Wheeler absorber theory of electromagnetism (Wheeler and Feynman, 1945; 1949), the modal interpretations (Dieks and Vermaas, 1998), the relational quantum mechanics of Rovelli (1996), which bases itself on observer-dependent states following Special Relativity but is non-committal about granting living or conscious status to the observers.

Prominent among the less-favored second category are the formulations by von Neumann (1955), supported and extended by London and Bauer (1983), Wheeler (1978) and Wigner (1961); the Many minds interpretation of Zeh (2000), Albert and Lower (1988); the work of Stapp (2001; 2009) in bringing out the key role played by the conscious observer in the measurement process through the mind-brain connection; the SQM formalism of Page (2011), and finally, the significant work of Manousakis (2007) in his consciousness-based interpretation.

With due apologies to the many other interpreters of QM, we do admit that the above is by no means an exhaustive survey of the landscape of interpretations proposed in either category. Some of the interpretations in the first category (*e.g.,* quantum logic) are fence-sitters and can easily accommodate the conscious observer. Some modify the formalism of quantum theory while others do not. The need to modify the QM formalism arises primarily because of the wish to keep the conscious observer out of the formalism, while the straight and simple fact is that it is inevitably present not only in QM but also in the entire scheme of science all the way right from the beginning.

We discuss this inevitable and undisputable role of the conscious observer in the scientific scheme in detail in the next section and then in section-3, the three main interpretational issues namely, the meaning of state, the measurement process and quantum non-locality are discussed. Section-4 gives the motivations for attempting a psychophysical interpretation of quantum mechanics and addresses each interpretational issue from a psychophysical standpoint. Section-5 discusses the intimate relationship of this interpretation with other major proposed interpretations. We conclude in section-6 with a discussion of quantum determinism.

## 2. The conscious observer in the scientific scheme

It is a fact that all our scientific theories are productions of very fertile brains of conscious observers and in order to make contact with the physical reality represented by observed phenomena, they must take in, and finally tally with, the observations of such conscious observers. Therefore, we discuss below this all-important, but usually played down in the name of objectivity and observer-independence, role of the conscious observer in the scientific scheme in general.

Indeed, *it is the conscious observer alone that gathers the data; classifies, organizes, analyses and interprets the data by looking at structural symmetries, regularities, periodicities etc; proposes hypotheses, postulates, laws and principles etc. which 'satisfactorily' explain these symmetries etc., often using mathematical tools; tests these hypotheses etc. by purposefully designing further fact-finding or fault-finding experiments; and then, if need be, enunciates new hypotheses etc. on the basis of more detailed and more refined data.* This is precisely the scientific method as has been in practice for the last few centuries. But, all the while, the practice and the claim has been to apply the method in as objective (i.e., observer-independent) a manner as possible keeping the all-important and ubiquitous

observer-- the data - gatherer, the data-interpreter, the hypothesizer, the fact-finder and also the fault-finder-- out of the scheme!

Evidently, there is a limit to this artificial isolation of the *observer* from the *observed* and our classical objective science program fails when we have to describe:

- properties (e.g., conjugate observables) of a microscopic system which gets inevitably disturbed by the very act of observation so that the accurate measurement of those properties becomes practically impossible (uncertainty principle).
- properties which depend on the mode of subject-object interaction in which case we cannot ascribe the properties to the system alone (complementarity).
- properties which are not purely objective i.e., the *felt qualia* which are dependent on the perceptions of the observing subject, e.g., colour of an object.

The first case is applicable even to purely classical measurements on macroscopic systems, but because of the ignorable smallness of the errors introduced in the measurement compared to the large values of the classical observables we conventionally ascribe the quantity to have the measured value, of course with the errors specified. Strictly speaking, as noted by Dirac (1947), we can never ever make an error-free measurement of the exact value of any quantity in practice, whether the system is classical or quantum mechanical.

The second and the third cases are also applicable in classical as well as in quantum mechanics always but we conventionally disregard them by resorting to an objectivity which is more a result of "*practical agreement among observers*" rather than an "*actual non-dependence on observers*". Scientific criteria like repeatability and verifiability etc. ensure that the last two cases are forcibly kept out till such time as they make their presence very strongly felt thereby forcing a paradigm shift in science. This has happened in case of quantum mechanics.

In the absence of any compelling scientific theoretical or experimental evidence for the 'material origins of mind' or for the 'mental origins of matter', and in view of psychophysical parallelism (Page, 2011; Pauli and Jung, 2001; Schrödinger, 1954; Schrödinger, 1967), we shall adopt the dualistic view in this work-- *The nature of Reality as a whole is neither fully objective nor fully subjective but is the result of the coming together of the subject and the object through the process of subject-object interaction*. The perceiving subject as *the observer*, the perceived object as *the observed* and the process of perception as *the observation*- all three come together to make up Reality. However, a series of rather *artificial* bifurcations are introduced in the process *by the observer* (Figure 1):

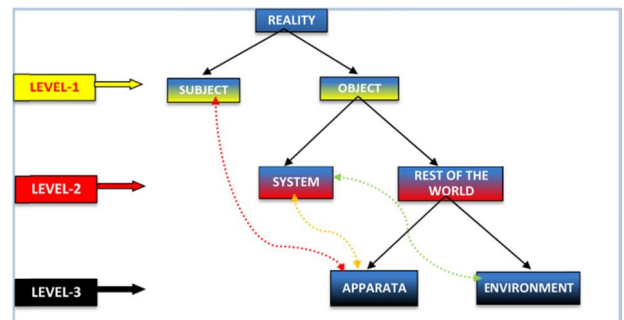

**Figure 1.** The three bifurcation levels employed in the scientific approach to *Reality*.

(a) She egoistically separates 'herself' out as the 'observing subject'- distinct and different from the 'observed object (the world)';

(b) She bifurcates the observed world into a 'system' and 'rest of the world' with the latter including her physical body also, unless the body itself is the system;

(c) She further partitions 'the rest of the world' into 'the apparata' (physical body included) and 'the environment', for scientific study;

And, after having done all this, *she also uses her human intellect for the analysis and inference but feigns subjective non-interference and vehemently claims that Reality is fully objective*!

This scenario is depicted in the fig-1 above where the dashed line connecting the system and the apparata represents the *interaction* between them during measurement while that connecting the subject and the apparata represents the *observation* of the result which completes the measurement process. The dashed line from the system to the environment represents the fact that no system can be absolutely isolated from the environment. Although this is a dualistic scheme, it is the most pragmatic one and in the present stage of development of our empirical sciences, it is the most acceptable and perfectly unbiased scheme regarding the nature of Reality, since we don't have any viable theory for consciousness or the mind.

## 3. Main Issues of Interpretation
### 3.1. The nature of the quantum state

What does the quantum state $|\Psi\rangle$ of a system represent? Three possible answers have emerged: (a) The objective state of the single system, (b) The subjective knowledge of the state of the single system, and, (c) An ensemble of systems.

Needless to say, even though the above three possibilities appear very distinct, all interpretations often have to struggle hard to interpret the wave function in a consistent way when it comes to measurement and verification of the probabilistic contents of the state. However, each proposed interpretation has its own advantages and oddities, and we need to develop a comprehensive interpretation which will encompass all the satisfactory aspects of the various interpretations proposed so far without any bias or prejudice towards a particular viewpoint. Further, we accept the view that as a matter of principle, *an interpretation should not modify the theory but should only interpret its formalism and establish thereby the correspondence with Reality which in this case is a consequence of the inevitable mutual interaction between the subject and the object*.

One rather surprising aspect of the proposed interpretations is that almost all of them fail to say even a word about the significance of the conjugate wave function $\Psi^*(\mathbf{r},t) = \langle\Psi|\mathbf{r},t\rangle$, the only exception being Cramer's transactional interpretation. Without the conjugate quantities no quantum mechanics is possible and yet they don't receive any straightforward interpretation and are treated as mere mathematical counterparts of the quantities concerned. This situation is sought to be remedied in the comprehensive interpretation proposed here. We will interpret here both $\Psi(\mathbf{r},t)$ and $\Psi^*(\mathbf{r},t)$ as being equally significant for the comprehension of QM.

### 3.2. Quantum measurement and state collapse

The second major point of interest is the issue of measurements in QM. A measurement leads to a specific eigenvalue for the measured observable from amongst a spectrum of possibilities. This collapse of the state to one *eigenstate* is a non-unitary process not describable by the Schrödinger equation which describes the unitary evolution of the state and thus becomes an independent postulate of QM. This also makes QM indeterministic and observer-dependent. Here, we encounter the debate concerning role of the conscious observer in the state collapse. According to von Neumann (von Neumann, 1955) and Wigner (Wigner, 1961) the chain of events in the measurement process logically culminates when the knowledge of the definite eigenvalue is registered in the mind of the observer. Without bringing in the conscious observer as the end-point of the causal chain we cannot escape the infinite regress.

The uneasiness in accepting the conscious observer as having a role (curiosity kills the cat!) is understandable since '*the abstract ego*' or the 'observer' in the von Neumann chain remains forever outside the formulation. It is entrusted with the job of collapsing the wave function in a non-unitary manner which is not describable by the Schrödinger equation. The main reasons for the uneasiness in accepting von Neumann's original proposal are:

- Firstly, we do not as yet know or have any scientific theory or formulation worth the name which can adequately describe the interaction of a conscious observer with the inanimate instruments, let alone the case of interaction with other conscious observers.
- Secondly, none of the original formulations of QM by Schrödinger, Heisenberg, Dirac or Feynman had any such intentions of describing the process of *conscious perception* and therefore QM is not expected to describe the same, which has mostly been in the domain of psychology, philosophy and, at best, of neurophysiology as far as scientific acceptability is concerned.
- Lastly, because of our classical reductionist training, we find it very

convenient to take shelter under the misty clouds of '*epiphenomenon*' or '*emergent phenomenon*' or '*complexity and self-organization*' when it comes to anything related to consciousness, its ramifications or interactions.

This shows that we, as conscious entities ourselves are really terribly afraid of ourselves and have failed miserably in dealing with ourselves i.e. with the conscious observers! Further, this state of affairs has stalled any real progress that we could have achieved by now in the last nearly one hundred years of struggling with the interpretation of QM by following an unbiased approach to possible observer participation along with the other purely objective approaches. Of course, the laudable approaches of Penrose-Hameroff (Hameroff and Penrose, 1996a, 1996b) and Stapp (Stapp, 2011) have been there for quite some time now and they aim at finding a mechanism of state collapse taking the observer's consciousness, or to be precise, the neurophysiological brain processes into account, but the central issue of the mind-brain connection, "the hard problem" of Chalmers (Chalmers, 196), still remains unsolved.

In the present work, we will attempt to give a detailed account of the measurement process including the conscious perception applying the QM formalism without any modification.

### 3.3. Quantum Non-locality and related paradoxes

Einstein, Podolski and Rosen (Einstein, Podolski and Rosen, 1935) gave a definition of realism thus: '*If, without in any way disturbing a system, we can predict with certainty (i.e., with probability equal to unity) the value of a physical quantity, then there exists an element of physical reality corresponding to this physical quantity.*' Following the maximal of signal propagation velocity as the light velocity, this criterion of locality just states the fact that '*there can be no communication between space-like separated points*'. What EPR showed was that if QM is correct, and if the above definition of physical reality is accepted as valid then locality cannot hold good, i.e., non-local correlations (spooky action-at-a-distance) must exist between space-like separated points.

One readily sees that the EPR definition of reality is valid only classically. This is because, the prediction unavoidably requires (a) the previous knowledge of some property of the composite system and (b) measurement on one component of the space-like separated system. However, Bell (Bell, 1964; 1966) showed and it has been experimentally verified (Aspect, Dallibard and Roger, 1982; Aspect, Grangier and Roger, 1982) that QM is correct and that non-locality and entanglement are an inherent fact of nature. However, considering that the EPR effect requires previous knowledge, measurement and subsequent inference by an observer, it does lend support to Heisenberg's 'knowledge interpretation' of the wave function and this is central to the present interpretation. We quote Heisenberg (Heisenberg, 1958):

> "*Therefore, the transition from the 'possible' to the 'actual' takes place during the act of observation…. We may say that the transition from the 'possible' to the 'actual' takes place as soon as the interaction of the object with the measuring device, and thereby with the rest of the world, has come into play; it is not connected with the act of registration of the result by the mind of the observer. The discontinuous change in the probability function, however, takes place with the act of registration, because it is the discontinuous change of our knowledge in the instant of registration that has its image in the discontinuous change of the probability function.*"

Heisenberg very clearly points to a psychophysical parallelism involved in the process of measurement with the observation of the result leading to the completion of the process of acquiring knowledge about the state, lacking which we were forced to admit of a probabilistic description of the pre-measurement state in terms a superposition.

### 4. The way out: Psychophysical Interpretation

As pointed out by von Neumann and Wigner, the conscious observer's subjective perceptions have to be taken into account if we are to have a complete quantum description of Reality. Attempts to achieve this goal have been mostly in the direction of finding out some underlying neurophysiological brain processes which collapse the quantum state in a measurement. However, it is quite surprising to discover that the all-powerful formalism of QM is already having all the requisites for a description of the psychophysical parallelism (Atmanspacher and Primas, 2006) without the need for any modification, and surprisingly enough, we have all along been very familiarly working with it. Only a little re-interpretation of the quantum mechanical the formalism is required.

**Table 1.** Psychophysical Parallelism

| PHYSICAL ASPECTS | PSYCHIC COUNTERPARTS |
|---|---|
| Physical system | *Mental image* |
| State: $|\Psi\rangle$ | *Knowledge:* $\langle\Psi|$ |
| Wave function : $\Psi(r,t) = \langle r,t|\Psi\rangle$ | *Conjugate wave function:* $\Psi^*(\mathbf{r},t) = \langle\Psi|\mathbf{r},t\rangle$ |
| Statistical frequency for ensemble | *Probability for single system* |
| Superposition | *Indefinite knowledge* |
| Collapse | *Definite knowledge* |
| Overlap of states (inner product: $\langle\Psi|\phi\rangle$) | *Comparison of (knowledge of) images:* $\langle\phi|\Psi\rangle$ |
| Projection operator: $P = |\Psi\rangle\langle\Psi|$ | *Quest: Is the state $|\Psi\rangle$ or image $\langle\Psi|$ ?* |
| Density operator: $\rho = \Sigma n P_n |\Psi_n\rangle\langle\Psi_n|$ | *Quest: How many in which state?* |
| Entanglement | *Knowledge of initial sate of composite system* |
| Non-locality | *Inference from previous knowledge* |
| Reduction of density matrix | *Neglect of DOF of one component subsystem* |
| Forward time evolution (Causal) | *Backward time evolution(Retrocausal)* |
| Retarded wave signals | *Advanced wave signals ($t \to -t$, $\mathbf{p} \to -\mathbf{p}$)* |
| Information sequence from system to brain | *Knowledge sequence from brain to system* |
| Many-worlds interpretation | *Many-minds interpretation* |
| Offer wave of transaction | *Confirmation wave of transaction* |

The very basic purpose of all science is the explanation of various phenomena on the basis of the simplest and the smallest set of principles. As discussed earlier, we are spatio-temporally localized observers and we perceive such localized sections of the entire universe as are perceivable by our senses and the mind through the use of various instruments. Accordingly, we describe in empirical sciences only what we sensory perceive or mentally conceive. A 'system' is thus an essentially inseparable part of the universe, artificially separated out by us in the process either of sensory observation or of mental abstraction. Such a system we usually characterize by (a) its configuration (b) its properties and, may be, (c) its utility. Some of its properties are the dynamical variables, and out of these, the physical observables are those that have real measurable values which make them more suitable ones for the characterization of the system state.

Table 1 above gives us the psychophysical parallelism as is evident in the QM formalism, while Table 2 given below succinctly summarizes the meaning of the postulates of quantum theory from the psychophysical perspective.

For the purposes of the discussion here, it is very important to clearly distinguish between knowledge and information. We define '*information*' as '*ordered data*' and '*knowledge*' as '*meaningful information*'. Thus, *information*, as a *measure of order* (in the sense of symmetry properties which give rise to the conserved quantities or observables for state characterization), is a completely objective physical quantity, while *knowledge*, through the association of meaning with information, relates to the conscious observer and hence is subjective. Thus, with this identification, the ket $|\Psi\rangle$ truly contains information, but the meaning, in the sense of association of structural, functional or relational attributes for purposes of comprehension, is obtained only when this information is decoded and is obtained with the help of the bra $\langle\Psi|$.

**Table 2. Elements of Psychophysical Interpretation**

| TERMINOLOGY | OBJECTIVE | SUBJECTIVE |
|---|---|---|
| 1. State | KET VECTOR : $\|\Psi\rangle$ | BRA VECTOR : $\langle\Psi\|$ |
| 2. Dynamical Variable | OPERATOR : **A** | ADJOINT OPERATOR : $\mathbf{A}^\dagger$ |
| 3. Physical Observable | HERMITIAN OPERATOR | SAME OPERATOR: $\mathbf{A}^\dagger = \mathbf{A}$ |
| 4. Expectation Value | $\langle\mathbf{A}\rangle = \langle\Psi\|\mathbf{A}\|\Psi\rangle$ | $\langle\mathbf{A}^\dagger\rangle = \langle\Psi\|\mathbf{A}^\dagger\|\Psi\rangle$ |
| 5. Dynamics | $i\hbar(\partial/\partial t)\|\Psi\rangle = \mathbf{H}\|\Psi\rangle$ | $i\hbar(\partial/\partial t)\langle\Psi\| = -\mathbf{H}\langle\Psi\|$ |
| 6. Pre-Measurement State | LINEAR SUPERPOSITION: $\|\Psi\rangle = \Sigma C_n\|\Psi_n\rangle$ | INDEFINITE KNOWLEDGE: $\langle\Psi\| = \Sigma C_n^* \langle\Psi_n\|$ |
| 7. Post-Measurement State | EIGENSTATE: $\|\Psi_n\rangle$ | DEFINITE KNOWLEDGE: $\langle\Psi_n\|$ |
| 8. Probability Amplitude | PHYSICAL OVERLAP: $C_n = \langle\Psi_n\|\Psi\rangle$ | MENTAL COMPARISON: $C_n^* = \langle\Psi\|\Psi_n\rangle$ |
| 9. Probability | FREQUENCY: $P_n = \|C_n\|^2$ | PROBABILITY: $P_n = \|C_n\|^2$ |
| 10. Projection Operator | $\mathbf{P}_n = \|\Psi_n\rangle\langle\Psi_n\|$ | $\mathbf{P}_n = \|\Psi_n\rangle\langle\Psi_n\|$ |

It may be the probability of obtaining an eigenstate as a result of measurement or the expectation of an observable (Born Rule) or a transition probability (Fermi Golden Rule) or whatever it is, we have to take the help of the bra $\langle\Psi\|$ to get meaningful information hidden in the ket vector $\|\Psi\rangle$. This also points to the fundamental role of conscious observation in measurement for acquiring information about a system. We express these important identifications as follows:

i) Information = Data $\oplus$ Order

ii) Knowledge = Information $\oplus$ Meaning

iii) Measurement = Interaction $\oplus$ Observation

The interaction between the system and the apparata in the measurement process does lead to a collapse, but unless and until it is observed leading to the rise of definite knowledge, the measurement remains incomplete and no meaningful information i.e. knowledge is obtained. Information is transacted during the interaction part of the measurement, while knowledge is obtained only by the observation of the pointer states of the apparata.

### 4.1 Interpreting the wave function and its conjugate

The most vital ingredient of the psychophysical interpretation is the interpretation of the complex conjugate quantities as representing the psychic counterparts of the corresponding physical quantities. The reason is that since the conjugate complex pair (Z, Z*) with $Z = x+iy = (x^2+y^2)^{1/2}\exp(i\phi)$ and $Z^* = x-iy = (x^2+y^2)^{1/2}\exp(-i\phi)$ are related by $Re(Z)=Re(Z^*)$, $|Z| = |Z^*|$, $Im(Z^*) = -Im(Z)$, and happen to be reflections of each other about the real axis on the complex plane, each of them can represent equally effectively the same real (measured) value of a physical observable. Thus, when the quantity $Z = |Z| \exp\{i(\mathbf{k}\cdot\mathbf{r} - \omega t)\}$ represents a retarded wave solution of any wave equation propagating fro past to future, $Z^* = |Z|\exp\{i(-\mathbf{k}\cdot\mathbf{r}+\omega t)\}$ represents the corresponding advanced wave solution propagating in the reverse direction from future to past. It is the observed causal sequence of emission and propagation (or absorption) events that leads us to discard the latter and retain the former only as representing physically meaningful solutions. Complex conjugation in this case is equivalent to the time-reversal and momentum-reversal transformations: $\{t \to -t, \mathbf{k} \to -\mathbf{k}\}$. We list below several clues to this identification of the complex conjugate as representing the mental counterpart of a physical phenomenon which is central to the psychophysical interpretation.

### (a) Indications from the actual mechanism of visual perception

As delineated above, advanced waves are the vehicles of mental perception through backward ray-tracing along a straight line of the signals received by the brain through the knowledge sequence. For every retarded solution for the wave function Ψ, we have a corresponding advanced solution represented by the complex conjugate wave function Ψ*. The psychophysical parallelism is most exactly represented if the external object denoted by Ψ has its mental image denoted by Ψ*. Further, backward tracing is essential for a causal comprehension of phenomena, since we can remember only the past and using this faculty of memory we can mentally go backwards into the past to verify the actual forward movement of the retarded wave solution as required by causality. In fact, it is actually the other way around: Causality is a consequence of our ability to remember only the past and not the future.

For the specific case of visual perception, for any object that is seen, there is a retinal image formed, but we don't 'see' this retinal image; instead, we see the object outside! How? This most basic phenomenon begs a proper scientific explanation. We propose that the advanced waves can greatly help us in this mater. The image cast by the object on the retina is truly left-right and up-down inverted as well as diminished in size. Similarly, an object moving east casts an image that moves west; a clockwise rotation casts a counter-clockwise rotating image and so on and so forth. *The mental reconstruction of the object as well as determination of its location in space-time giving rise to determinate perception is impossible to comprehend unless backward ray-tracing by the conscious mind using advanced waves is taken recourse to.*

To understand the role of advanced waves in visual perception the case of perception of virtual images is very illuminating. The traditional optics textbook statements right from the days of Euclid through the times of Kepler (Lindberg, 1986) and Newton (Newton, 1952) till today like - "*rays appear to come from the point where the virtual image is formed*"- need to be probed further regarding how such '*appearances*' come about. If there is a physical object, or more generally, a source of light- a point wherefrom real retarded-wave light rays emerge or where they meet, then the perception of the said object or point is easily explained. But, when there is neither of them as in the case of a virtual image, how can the corresponding visual perception come about, unless the retarded wave light rays are mentally retraced backwards in space and time? And, this is precisely what the advanced waves achieve.

An interesting example of the independent existence of these advanced waves would be that of the perception of an atmospheric mirage as depicted in fig. 2 below. The upright object BC is 'seen' by the observer O to have the inverted image BD on account of a gradual decrease with height of the refractive index owing to the temperature gradient of air. The physical retarded signal from C to O (green line) has its mental advanced wave counterpart from O to C (dashed brown line). When the retarded signal from C via the curved refraction path CAO is mentally reversed, the mind retraces along the straight line OA up to A, and further on along the straight line path from A to D, and not along the reversed curved path from A to C, although advanced waves are present alongside the retarded waves in every segment in this curved path. The mental retracing via the advanced waves follows the straight line path from A to D, as if absolutely unaffected by the presence of air, thus giving the perception of the inverted image BD.

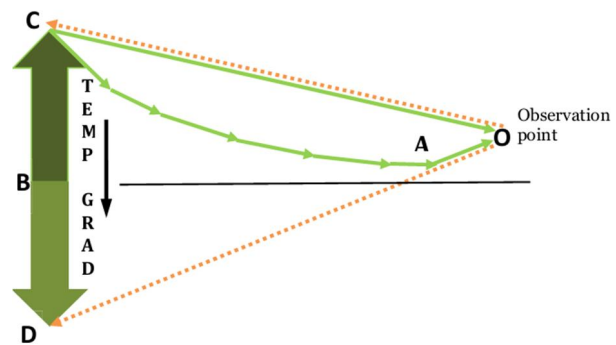

**Figure 2.** Atmospheric mirage as an illustration of an advanced wave phenomenon.

This also shows that for matter-matter interactions, the familiar Maxwellian retarded-wave electrodynamics is sufficient and there is no need for bringing in advanced waves. It is only when we want to describe conscious perceptions by mind-matter interaction (e.g. virtual images) that advanced waves are required. Basing on this, one may also speculate about explaining the ability of the mind to have independent perceptions via the past-directed advanced waves whether the corresponding physical retarded waves and their sources exist or not, as in the case of recollection of the past from memory or in the case of dream perceptions. Indeed, taking the mind as a perceiver of advanced waves through the physical brain which interacts with the external world through the combination of advanced and retarded waves.

To further shed light on the observer-participation in mirage perception, we may ask *a la* Einstein (Mermin, 1985): "*Is the mirage out there when nobody looks?*" Clearly, it's the backwards tracing by the observer's mind that is responsible for the mirage. We note that a similar phenomenon also happens in case of auditory perception, when, for example, the sound waves from the source come to the observer through a detour and by mental backward retracing the mind locates the source to be situated in the direction of reception rather than its actual location.

### (b) Indications from Wheeler-Feynman absorber theory

The afore-mentioned one-to-one correspondence between the advanced and the retarded solutions of Maxwell's equations has been exploited by Wheeler and Feynman (*ibid*.) to propose the absorber theory of electrodynamics. According to their proposal, the time-symmetric combinations of half-advanced and half-retarded waves emanating from the emitter as well as from a future absorber can lead to the same consequences as we find with purely retarded waves emitted by the emitter in conventional Maxwellian Electrodynamics, the only difference being the fact of the participation of the absorber. The question is: why do we find all EM phenomena taking place with the perfectly causality-respecting retarded waves only and why there is no direct experimental proof of the existence of Retrocausal advanced waves?

It is clear that matter-matter interactions can be understood completely in terms of retarded waves but if the absorber happens to be the physical sense organ of a conscious observer, then the very process of causal reconstruction described above requires the mental reversal of time and momentum of the retarded signals, which are nothing but the corresponding advanced waves. Thus, it might be postulated that Nature does use the advanced waves in the process of acquiring knowledge by a conscious observer. Feynman (Feynman, Leighton and Sands, 1964a) says:

"*Now, one problem is, by what process do we see light? There have been many theories, but it finally settled down to one, which is that there is something which enters the eye- which bounces off objects into the eye. We have heard that idea so long that we accept it, and it is almost impossible for us to realize that very intelligent men* (obviously referring to himself and Wheeler in his own inimitable way) *have proposed contrary theories- that something comes out of the eye and feels for the object, for example.*"

The Wheeler-Feynman theory would thus perfectly well account for the mind-matter interaction, with the mind utilizing, in the process of observation, the advanced waves to gain knowledge of the matter-matter retarded wave interactions.

One may speculate about purely mind-mind interactions via advanced waves only, which would explain a whole lot of accumulated data on Retrocausal phenomena not strictly describable by our sciences so far. It is then most natural to conjecture that the Psychophysically Extended Wheeler-Feynman theory would be as follows:

(a) Matter-matter interaction— Purely Retarded waves

(b) Mind-matter interaction— Half Retarded waves +Half Advanced waves

(c) Mind-mind interaction— Purely Advanced waves (see the discussion in section-6)

The psychophysical parallelism thus leads us quite naturally to identify $\langle\Psi|$ with knowledge of the state $|\Psi\rangle$. It further reveals that electromagnetic signals have another dual character: They can couple to both, matter as well as mind, via the retarded and advanced waves respectively. What ordinarily concern us in physics are the purely matter-matter interactions which are quite well explained by only the retarded waves of classical electrodynamics. And, It is only when we move on to quantum theory to describe Quantum measurements with the apparata-observer interaction leading to rise of knowledge and the consequent state collapse, we need the Feynman-Wheeler formalism of half advanced and half retarded waves.

### (c) Indications from the transactional interpretation

The Wheeler-Feynman theory was used by Cramer (*ibid*) in the Transactional interpretation (TI) to interpret the solutions $\Psi^*(\mathbf{r},t)$ of the conjugate Schrödinger equation:

$i\hbar \partial \Psi^*/\partial t = - \mathbf{H}\Psi^*$

as the advanced waves of the Wheeler-Feynman absorber theory, which are emitted by the absorber to enable the confirmation of a transaction with the emitter, thereby establishing the Born rule $P=\Psi^*\Psi$ for probability of confirmation at the emitter location.

In TI, the confirmed transaction is equivalent to the collapse of the Copenhagen Interpretation and it is proclaimed that there is no need for a conscious observer and that only the *material absorber* is required (Cramer, 1986). But, it does not provide the answer to the question as to which one among all the possible transactions is materialized in a particular measurement unless and until the confirmed transaction (collapsed state) is confirmed again (i.e. known) by the conscious observer! A confirmed transaction is thus the result of two back-to-back transactions: the (system-apparata) interaction transaction followed by the (apparata-observer) observation transaction which leads to the rise of definite knowledge and thereby collapses the pre-measurement state. This can be achieved only by the use of the advanced waves in the second transaction between the conscious observer and the absorber (detector or apparata). Thus, it is clear that the TI cannot avoid the conscious observer as the 'ultimate collapser' of the state.

### (d) Indications from the ABL Time-Symmetric Formalism (TSF)

The past-directed bra vectors have been used on an equal footing with the future-directed ket vectors as the essential ingredients in a time-symmetric formalism (TSF) of quantum mechanics developed by Aharonov (Aharonov *et al.,* 1964). Though this formulation yields results completely in agreement with the standard formulation, the state of a quantum system is not completely specified by the ket vector only but by conjoining it with a bra vector. In particular, any ideal von Neumann measurement at time t that collapses the system to the future-directed ket vector $|\Psi(t')\rangle$ for $t'>t$, also creates simultaneously the past-directed state $\langle\Phi(t'')|$ with $t''<t$ so that the state of the system within the relevant time interval between two successive measurements is completely specified by the two-state vector $\langle\Phi|\ |\Psi\rangle$. However, the time symmetry envisaged in the TSF is not automatically ensured for all situations since the past is certain while the future is not. However, at the exact instant of measurement which creates these states we have:

$$|\Psi(t)\rangle = \lim_{t'\to t} (\ |\Psi(t')\rangle) = \{\lim_{t''\to t} (\langle\Phi(t'')|)\}^{\dagger} \quad (1)$$

Due to the finite propagation speed of the signal from the apparata, the observer has got to extrapolate mentally into the past to get to the state at the measurement instant t in order to infer about the state of the system at that moment from his current brain state. Indeed, the TSF is a direct representation of psychophysical parallelism and the entire QM formalism can be recast and reinterpreted in terms of the bra-ket as representing the state rather than the single ket with the bra left uninterpreted. The back-ward evolving bra however can have its interpretation only as representing a mental state since it is the mind alone (apart from antiparticles in the Feynman-Stuckelberg interpretation) that has the peculiar ability to move backwards in time. We are thus led in a rather straightforward manner by the TSF to interpret the past-directed bra as the knowledge state or the mental counterpart of the future-directed physical state represented by the ket vector as proposed here.

### (e) Indications from the dual nature of probabilities

It is well known that Probabilities possess a dual character: objective (frequency interpretation) as well as subjective (belief interpretation). Mainly, the Bayesian conditional probability approach has been in the forefront of all research aimed at incorporating the dual aspects of probabilities in quantum theory. However, a very simple and straightforward non-Bayesian approach has been proposed by the author (Pradhan, 2011) in a recent work. The quantum probability given by the Born rule $P = \Psi^*\Psi$ suggests that we can straightaway take the conjugate wave function Ψ* to represent the mental counterpart of the physical wave function Ψ, such that the dual nature becomes self-evident.

If we accept a dualistic view of reality, then all these indications from various perspectives do put forth a very strong case for interpreting the bra as the mental state corresponding to the ket as the physical state. The psychophysical interpretation then explains the emergence of a real physical world from the probabilistic quantum world as arising from the interaction of the conscious observer with the latter. The mental states of the observer are the past-directed bras which are compared with what is received by the brain through the sensory apparata in a measurement, and accordingly the system properties are ascertained. For the purposes of prediction of outcome of measurements the subjective mental amplitude Ψ* is multiplied by the objective physical amplitude Ψ as it should be for independent probabilities, and this explains the Born rule.

More explicitly, with reference to table-II above, for a given system described by the general superposition state $|\Psi\rangle = \Sigma C_n |\Psi_n\rangle$, the objective amplitude for collapse to state $|\Psi_n\rangle$ is the overlap function $C_n = \langle\Psi_n|\Psi\rangle$, while the corresponding subjective amplitude for the same is given by the comparison function $C_n^* = \langle\Psi|\Psi_n\rangle$ representing the knowledge of the objective amplitude $C_n$ and the probability is then given by the product of these two amplitudes $P_n = C_n^* C_n = \langle\Psi|\Psi_n\rangle \langle\Psi_n|\Psi\rangle = |C_n|^2$, which is the Born rule.

All the above different approaches suggest that we can interpret the formalism of quantum mechanics in a complete manner only by interpreting the bra vector and the conjugate wave function as the mental counterpart of the physical ket vector and the corresponding wave function respectively.

## *4.2. The measurement problem and state collapse by the conscious observer*

In general, we must agree upon the fact that a measurement is completed only upon the observation of the results and not before that. Otherwise, we have "measurements with unknown results" or "unobserved measurements" which serve no meaningful purpose whatsoever. Even when the results in an experiment are null or inconclusive, such nullity or inconclusiveness must be *known* by an observation of the apparata.

Thus, if the system-apparata (matter-matter) interaction constitutes the objective half of the measurement process, then the apparata-observer (matter-mind) interaction leading to the knowledge of the system state may be said to be the subjective half.

The conscious observer (subject) here may be identified with the 'abstract ego' of von Neumann, since the entire physical universe including the physical body of the observer is in the object part of the bifurcation level-1 of section-2 above. As mentioned in the previous section, while doing science objectively, we do always strive to keep ourselves, i.e. the observers, out of the scheme, which naturally presupposes a non-physical or, at the very least, a non-material (hence called 'abstract' by von Neumann) subjective ego that actually cognizes everything.

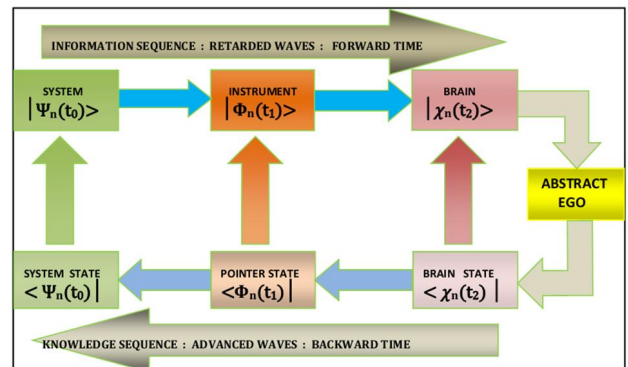

**Figure 3.** Schematic of an actual von Neumann measurement process.

The process of observation is then a perception of the result of the system-apparata interaction through the apparata-observer interaction, where, as mentioned earlier, the apparata include the measuring instruments and also the physical body (senses and the brain etc.) of the observing subject. The actual process of a measurement leading to the rise of definite knowledge of these observables proceeds by the following two sequences.

## (a) Information Sequence (IS): The Physical part

IS-1: Electromagnetic signals from the system (or apparata) to the sense organs,

IS-2: Nervous signals from the senses to the brain,

IS-3: Excitation of a corresponding brain state or neural correlate (NC).

As usual, these steps do proceed by the familiar retarded waves in the forward time direction. But, we get the knowledge by drawing inference through establishing a causal linkage from the effects to the cause in a backward time direction by mentally reversing the time and the momenta in all processes involved in the perception (fig. 3).

It is common experience that our minds are endowed with the uncanny ability to move backwards in time by reversing the information sequence. The ego focuses the attention on the NC and identifies itself with it and then starts the process of time and momentum reversal of all the signals that led to the formation of the NC. From the interface of the signals with the sense organs such as the eyes, a momentum reversal along a straight line path from the reception point is mentally done backwards in time and results in the perception of the object outside. The cognition (definite knowledge of state) of the system or object then takes place by comparison with previously stored images in the memory. Thus the following *knowledge sequence* results:

### (b) Knowledge Sequence (KS): The Psychic part

KS-1: Mental reversal of the excitation of the NC forming the brain state backwards in time.

KS-2: Mental reversal of the nerve currents (from the senses to the brain) backwards in time.

KS-3: Mental reversal of the (electromagnetic) signals received by the senses backwards in time.

The steps of the Knowledge Sequence take place with the help of the advanced waves propagating in the backward time direction from the observer. Usually, in the last step KS-3, the mental reversal occurs along a straight line from the sense organ outwards, even if the original incoming retarded waves might have suffered reflections or come along curved paths, as evidenced by the perception of virtual images and mirages etc. It is rather surprising to note that the steps in the Knowledge sequence as well as the comparison required for the rise of definite knowledge are already adequately described by the quantum formalism as tabulated in the above two tables.

To briefly explain with reference to fig-3 above, we note that what is actually observed in any act of observation is the brain state or neural correlate $|\chi_n(t_2)\rangle$ resulting from the signals received through the sense organs and conveyed through the nerve channels to the brain. The lack of definiteness before the observation vanishes upon the registering of this definite state as an element of knowledge. This brain state has the corresponding pointer state $|\Phi_n(t_1)\rangle$ as its source and in its turn the pointer state comes about because of interaction with the system in state $|\Psi_n(t_0)\rangle$. The forward time sequence in the information reception process in the objective half is through retarded signals and therefore there must be different successive times $t_0$, $t_1$ and $t_2$ of the events at the system, instrument and brain respectively. Similarly, the processing in the brain for generation of knowledge must proceed via backward propagation (or mental reversal) of the signal reception sequence. This is the reason why we usually have the sense of "now" in regard to any perception irrespective of the distance of the object perceived.

As in Cramer's transactional interpretation, we take the advanced signals to be the momentum-reversed and time-reversed counterparts of the retarded signals. However in our interpretation, *while matter-matter interactions proceed through retarded signals, in case of mind-matter interactions resulting in knowledge, the ego sends forth advanced signals to interpret the information received and thereby gains knowledge.*

First of all, the ego reverses the neural excitations which formed the correlate corresponding to $|\chi_n(t_2)\rangle$ and thereby gains the knowledge of the definite brain state; then it moves backwards through the advanced nerve signals to the retinal image and then through the advanced electromagnetic waves it moves further backwards to reach the instrument at time $t_1$ which collapses to a state of definite knowledge $|\Phi_n(t_1)\rangle$ upon the gaining of this knowledge; and then moves on through the advanced signals to the system to reach it at time $t_0$ and collapses it to the state $|\Psi_n(t_0)\rangle$ upon gaining such knowledge. The association $|\chi_n(t_2)\rangle \Leftrightarrow |\Phi_n(t_1)\rangle \Leftrightarrow |\Psi_n(t_0)\rangle$ by the ego is done in such a manner that the observer presumes that the pointer or the

system is observed now i.e. at time $t_2$, the moment of the observation of the brain state, when the measurement process gets completed.

We now derive the Born rule in the psychophysical interpretation. Consider a detailed POVM in a complete von Neumann measurement with the Projection operators

$$\mathbf{P}_n = |\Psi_n\rangle\langle\Psi_n| \qquad (2)$$

as the Krauss Operators. The premeasurement state of the system is given by the superposition

$$|\Psi\rangle = \Sigma \langle\Psi_n|\Psi\rangle|\Psi_n\rangle \qquad (3)$$

which, upon measurement, evolves to the state post-measurement unnormalised state

$$\mathbf{P}_n|\Psi\rangle = |\Psi_n\rangle\langle\Psi_n|\Psi\rangle = \langle\Psi_n|\Psi\rangle|\Psi_n\rangle = C_n|\Psi_n\rangle \qquad (4)$$

For the rise of definite knowledge of the post-measurement state of the system, the observer has to compare the resulting state $\mathbf{P}_n|\Psi\rangle$ with the corresponding copies $\langle\Psi_m|$ previously stored in the memory, and then update his knowledge accordingly by replacing $|\Psi\rangle$ with $|\Psi_n\rangle$ as the new state of the system. The result of this comparison

$$\langle\Psi_m|\mathbf{P}_n|\Psi\rangle = \langle\Psi_m|\Psi_n\rangle\langle\Psi_n|\Psi\rangle = \delta_{mn}\langle\Psi_n|\Psi\rangle = C_n \qquad (5)$$

is non-vanishing only for the projected state due to the orthonormality of the basis states, which are mutually non-overlapping distinct eigenstates

$$\langle\Psi_m|\Psi_n\rangle = \delta_{mn}. \qquad (6)$$

When the system was assumed to be in $|\Psi\rangle$ before measurement, it had unit norm: $\langle\Psi|\Psi\rangle = 1$. Now, after the measurement, the norm of the new state $\mathbf{P}_n|\Psi\rangle$ will give us the measure of "How much of $|\Psi\rangle$ was along $|\Psi_n\rangle$?", which is the probability of obtaining $|\Psi_n\rangle$:

$$P_n = \langle\Psi|\mathbf{P}^\dagger_n\mathbf{P}_n|\Psi\rangle = \langle\Psi|\mathbf{P}_n|\Psi\rangle = \langle\Psi|\Psi_n\rangle\langle\Psi_n|\Psi\rangle = |C_n|^2 \qquad (7)$$

Please note that the whole analysis can be done equally well with the pointer states $|\Phi_n\rangle$ of the apparata or with the brain states $|\chi_n\rangle$ of the observer because of the entanglement which guarantees one-to-one correspondence of all the three sets of states.

One important theme of the interpretation proposed here is the fact that any measurement must involve observation alongside interaction and that the brain state corresponding to the system state $|\Psi\rangle$ is nothing but the image state $\langle\Psi|$. Thus, there are two collapses: the objective one is the *interactive collapse* and the subjective one is the *observational collapse* that completes the measurement. Between the instant of interaction ($t_1$) and the moment of observation ($t_2$) by an observer, the system and the apparata remain entangled for all observers while the subjective collapse occurs only for the observer(s) that make(s) the observation. For all the other observers, they either have to make fresh observations on the apparata for themselves, or, have to *believe* in the account given by the first observer, in order for subjective collapse to occur in them regarding the outcome of that particular measurement.

### 4.3. Non-locality and related paradoxes

Quantum non-locality in all situations can be explained as arising from the observer's inference basing on previous knowledge of correlations between constituent subsystems of the whole system. The paradoxes cease to be paradoxes once the knowledge of the conscious observer is taken into account as an element of Reality. We consider below the most bizarre kind of non-local situation possible - A Universal entanglement, and show how the EPR like paradoxes are very simply explained.

Consider a system 'S' (which may, for example, be a Hydrogen atom) and the rest of the world $S' = A \cup \mathcal{E}$. Here, the apparata A include the brain of the observer and $\mathcal{E}$ is the environment such that the whole objective Universe $U = S \cup S' = S \cup A \cup \mathcal{E}$ is described by the direct product Hilbert space $H_U = H_S \otimes H_{S'} = H_S \otimes H_A \otimes H_\mathcal{E}$. As per the psychophysical interpretation, if, upon observation, the observer's brain state collapses to the state corresponding to the $m^{th}$ pointer state confirming the eigenvalue m, the system, the apparata and the environment all collapse simultaneously to the state m. The universal pre-measurement pure state is then described by the globally entangled state in the product space:

$$|\Omega\rangle = \Sigma_m C_m |\Psi_m\rangle|\Phi_m\rangle = \Sigma_m C_m |\Psi_m\rangle|A_m\rangle|\mathcal{E}_m\rangle, \quad \Sigma_m |C_m|^2 = 1 \qquad (8)$$

The reason for the entanglement being the fact that the quantum numbers 'm' are those that characterize globally conserved quantities like energy, charge, angular momentum etc. which follow from very general symmetry considerations in quantum theory. The corresponding universal pure state density matrix is given by:

$$\boldsymbol{\rho} = |\Omega\rangle\langle\Omega|$$
$$= \Sigma_{mn} C_m C^*_n |\Psi_m\rangle|\Phi_m\rangle\langle\Psi_n|\langle\Phi_n|$$
$$= \Sigma_{mn} C_m C^*_n |\Psi_m\rangle|A_m\rangle|\mathcal{E}_m\rangle\langle\Psi_n|\langle A_n|\langle\mathcal{E}_n| \qquad (9)$$

Now suppose, an observer makes a measurement of the energy of the system S with the help of apparata A, and as usual, we have to ignore the vast no of DOF of the rest of the world which cannot be taken care of in any conceivable way. As is well known from the environment-induced decoherence effects, *einselection* (Zeh, 2010; Paz and Zurek, 1999) singles out the energy eigenbasis as the preferred pointer basis for the rest of the world $S' = A \cup \mathcal{E}$ composed of the apparata A and the environment $\mathcal{E}$. The environment states $|\mathcal{E}_n\rangle$ thus also quickly become orthonormal, i.e., $\langle\mathcal{E}_m|\mathcal{E}_n\rangle \to \delta_{mn}$. This means that by tracing over the states of $\mathcal{E}$, the reduced density matrix $\boldsymbol{\rho}_{sA}$ for the system S and apparata A will be of the form:

$$\boldsymbol{\rho}_{sA} = tr_{\mathcal{E}} \boldsymbol{\rho} = \Sigma_{mn} \delta_{mn} C_m C^*_n |\Psi_m\rangle|A_m\rangle\langle\Psi_n|\langle A_n|$$
$$\to \Sigma_n |C_n|^2 |\Psi_n\rangle\langle A_n|\langle\Psi_n|\langle A_n|$$
$$= \Sigma_n |C_n|^2 \mathbf{P}_n^s \otimes \mathbf{P}_n^A \qquad (10)$$

where, $\mathbf{P}_n^s$ and $\mathbf{P}_n^A$ are respectively the projectors onto the eigenstates of the system and the apparata (observer).

But, the total energy of the whole universe is a globally conserved quantity having value, say, $E_o$. Now, making a measurement of energy on S, if one obtains a value $E_n$ for the system which has a probability $|C_n|^2$, then from previous knowledge of the total energy for the whole universe, one immediately gets the energy of the rest of the world i.e. $S' = A \cup \mathcal{E}$ as:

$$E_n' = E_o - E_n \qquad (11)$$

The apparatus, thanks to decoherence, acts in this case as a simultaneous measurer of the energy of S and S′ both. In fact, it is highly plausible that the total energy of the whole universe is exactly zero (Bermann, 2009) i.e., $E_o=0$, in which case, $E_n'=-E_n$, and the pointer states precisely become the preferred basis states of both the system and the rest of the universe.

This means that a perfectly local measurement of the system energy by an observer has yielded the value of the energy of the rest of the world which obviously contains vast regions of space, which are space-like separated from the system. Now, what does this mean? Has any Einsteinian *spooky-action-at-a-distance* taken place or is it merely the inference from previous knowledge?

It is clear that in all EPR-like situations it *is* the previous knowledge of the observer that leads to the correlatedness and there is nothing surprising in this. The paradox is resolved once the previous knowledge of the observer is taken into account through psychophysical parallelism. The detailed mechanism course has to be that of the transactional scheme as shown by Cramer (Cramer, 1980) but with the additional interpretation of the advanced waves as knowledge producing signals for the conscious observer. The conclusion here is the same as arrived at by Smerlak and Rovelli using the relational arguments (Smerlak and Rovelli, 2007), the only difference is that they *consciously* try to avoid granting *conscious* status to their *observers*.

## 5. Connection with other major interpretations

The psychophysical interpretation is in no conflict with many of the major interpretations proposed so far, rather it augments them by incorporating the conscious observer into the scheme as an equal partner in determining reality as a whole. Since it keeps intact the formalism of QM, it obeys all the tenets of the *Copenhagen* interpretation. It bases itself on Heisenberg's *knowledge* interpretation with the distinction that it is the bra that encodes the knowledge of the system which is represented by the ket. It supports the idea of state collapse by the abstract ego of the conscious observer as envisaged in the *von*

Neumann approach to quantum measurements. As such, any interpretation that keeps intact the formalism of QM will have no conflict with the psychophysical interpretation.

It is truly a *many-minds* interpretation and reduces to the *many-worlds* interpretation in the special case when one takes, as usual, the bra as merely the mathematical complex conjugate of the ket bereft of any mental significance. But, still then, it requires in addition the existence of at least one conscious observer (mind or psyche), for otherwise, if all the worlds are bereft of consciousness then the interpretation problem itself vaporizes and vanishes in toto? The many worlds, in order to make sense, do require the cognizance of at least one of them, like the one inhabited by us, by at least one conscious observer, which may of course be a cosmic observer, if not an earthling like us. It is of course a matter of future research to truly comprehend the relationship of such a cosmic consciousness with individual centers of consciousness inhabiting the component worlds in regard to the measurement problem (Pradhan, 2010).

It respects and heavily depends upon the *transactional* interpretation through the use of past-directed advanced waves for gaining knowledge of the system. It extends the *relational* interpretation by allowing for states relative to conscious observers. It augments *Manousakis' formulation* of QM (Manousakis, 2007) on the basis of conscious perceptions through the psychophysical parallelism. It keeps alive the hopes of finding the 'location' in the physical brain for the ego, the ultimate collapser of the state, as envisaged in the valiant attempts by Penrose-Hameroff (Hameroff and Penrose, 1996a) and Stapp (Stapp, 2011) employing '*interactive dualism*'. We remark in this connection that quite unlike many of the founding fathers of QM, most of us by routinely avoiding and shying away from matters relating to consciousness not only impoverish science but also do a great disservice to its claims of being an unbiased approach to Truth.

## 6. Discussion and Conclusion
The psychophysical interpretation is '*comprehensive*' for three reasons:

- It *comprehends* in its bosom the basic truths of many previous interpretations.
- It aids the *comprehension* of many of the subtleties of quantum theory.
- It is based on the analysis of the process of conscious perception or *comprehension*.

It makes a quantum measurement a two-step process. The state remains a superposition as long as the knowledge remains indefinite, even though the system-apparata interaction might have collapsed the system to one of the eigenstates. The cognition of the pointer state finally leads to definite knowledge and the knowledge state collapses. The pre-measurement physical state $|\Psi\rangle$ of the system matches with the pre-measurement knowledge state $\langle\Psi|$ i.e. $\langle\Psi|\Psi\rangle = 1$, but the post-interaction physical state say, $|\Psi_n\rangle$, does not match with the knowledge state $\langle\Psi|$, but has an overlap $C_n = \langle\Psi_n|\Psi\rangle \neq 1$, till the time of observation of the pointer state. The post-observation knowledge state $\langle\Psi_n|$ matches perfectly well with the post-interaction physical state $|\Psi_n\rangle$ i.e. $\langle\Psi_n|\Psi_n\rangle = 1$, and we have the collapse process completed with the rise of definite knowledge. Between the time of interaction and observation, the probabilistic description holds just as it did for the pre-measurement period. The great advantage is that this formulation applies equally well not only to classical and quantum measurements, but also to any ordinary process of conscious perception and cognition.

### Determinism and Free-will
This comprehensive Psychophysical interpretation opens a new vista for perfect determinism in QM through the use of advanced waves. The future can be predicted with certainty if the advanced waves from the future that are intercepted by the brain are captured and interpreted by the mind, which unfortunately is mostly preoccupied either with the job of interpreting the continuously impinging retarded wave signals through various senses or with the job of sending past-directed waves from the present to recall past events stored in the memory. Probably, the way out is to stop this routine manner of incessant conscious and subconscious activity of the mind, to calm it down to an almost thoughtless condition, so that it can be made receptive enough to capture and interpret the feeble advanced wave signals from the future events reaching the brain.

The traditional argument (Feynman *et al*, 1964b) that a future event F, if known in advance, may be prevented from materializing '*by doing the right thing at the right time*' i.e. by appropriate rearrangements made in the present set-up P, does not hold, since in that case it is the advanced wave signals from the 'altered future event' F′ due to the subsequently changed present P′ (including those from P′ itself which is in the future of P), that would be intercepted and accordingly the latter (*i.e.,* F′) would be predicted in the first place and not F. If F were really not to happen due to our re-arrangements, then obviously that could not be known as a future event. Now, this brings in the question of 'free will' also into the picture— can we not really make, by our own free choice, such willful changes as would really prevent F from occurring? The answer is *no*, because of the logical contradictions it engenders.

Further, when we meekly and readily accept without any arguments our inability to alter the past in any manner whatsoever, why should we complain if we are likewise denied any free-will to intervene in the future affairs because of time-symmetry? For example, we are forced to helplessly accept the future certainty of death of every one of our human society, and that too, certainly within a span of, say, a maximum of 125 years after the birth, and we can hardly do anything about it! Only the exact timing of the event is uncertain, but not the event itself! Or, to take less frightening examples, we can't change, by any amount of exercise of our so-called free-will, either the universal constants (like Planck's constant) or the many constraints imposed by them on us or the cosmic-scale phenomena like the expansion of the universe or, for that matter, even a solar eclipse. The concept of Free-will is truly of extremely restricted validity. The question of free-will arises as long as the future is uncertain. But, once the future is known with certainty by an individual subject, the free-will simply ceases to operate. In fact, it is the other way around: Perfect knowledge (quantum determinism) of the future arises only in one who has given up individual free-will.

Traditionally, the advanced waves are branded as unphysical and so also the mind. It may be conjectured that the mind is nothing but a dynamic centre of incoming (from future to present) and outgoing (from present to past) advanced waves around the Central Nervous System (CNS), always busy generating, receiving and interpreting them. The detailed mechanism of this knowledge-producing aspect of advanced waves is again a matter for serious future research.

Further, it should be made clear that neither the age-old Cartesian subject-object dualism nor the psychophysical parallelism may be an ultimate fact, and, the mind may finally prove to be having an upper hand over matter, purely because of the subtlety of the advanced waves, if not for anything else. But, for the time being we do not delve into this issue and leave it for future work. The more our mainstream researches are reoriented and focused towards scientific investigations of consciousness with an open-mindedness that should be characteristic of anything worth calling science, the more fascinating will be the fundamental discoveries that are sure to be made both in theory and experiments alike. The road from the widely accepted materialist-reductionist approach to the Ultimate Reality has to be waded through an humble appreciation of the Dualist view in its fullest ramifications as an interim measure, and then and only then can we confidently work towards a unified formalism of '*Reality as whole*' and a '*Truly Final Theory of Everything.*'

### Acknowledgements

The author is grateful to L. P. Singh for discussions. The ongoing technical support of Institute of Physics, Bhubaneswar is also gratefully acknowledged.